\documentclass{moriond}

\usepackage{wrapfig}
\def\Journal#1#2#3#4{{#1} {\bf #2}, #3 (#4)}

\def\NPA{{\em Nucl. Phys.} A}

\def\PRL{\em Phys. Rev. Lett.}
\def\PRD{{\em Phys. Rev.} D}

\def\ApJL{\em Astrophys. J. Lett.}
\def\ApJ{\em Astrophys. J.}
\def\PhR{\em Phys. Rep.}
\def\ARNPS{\em Ann. Rev. Nucl. Part. Sci.}
\def\CQG{\em Class. Quant. Grav.}

\def\be{\begin{equation}}
\def\ee{\end{equation}}
\def\bea{\begin{eqnarray}}
\def\eea{\end{eqnarray}}

\newcommand{\Photo}{}
\newcommand{\chimera}{{\sc Chimera}}
\newcommand{\msun}{\ensuremath{M_\odot\,}}
\newcommand{\etal}{~\emph{et al.}}

\begin{document}
\vspace*{4cm}
\title{GRAVITATIONAL WAVE SIGNALS FROM MULTI-DIMENSIONAL CORE-COLLAPSE SUPERNOVA EXPLOSION SIMULATIONS}

%

\author{K.N. Yakunin$^{a}$, E.~Endeve$^{a,b,d}$, A.~Mezzacappa$^{a,b}$,\\
        M.~Szczpanczyk$^{e}$, M.~Zanolin$^{e}$
        P.~Marronetti$^f$,\\ E.J.~Lentz$^{a,b,c, g}$,
        S.W.~Bruenn$^h$,
        W.R.~Hix$^{a, c}$,\\ O.E.B.~Messer$^{a, c, i}$,
        J.M.~Blondin$^{j}$, 
        and J.A.~Harris$^{i,k}$
        }

\footnote{$^{}$Department of Physics and Astronomy, University of Tennessee, Knoxville, TN 37996-1200, USA}
\footnote{$^{}$Joint Institute for Computational Sciences, Oak Ridge National Laboratory, P.O. Box 2008, Oak Ridge, TN 37831-6354, USA}
\footnote{$^{}$Physics Division, Oak Ridge National Laboratory, P.O. Box 2008, Oak Ridge, TN 37831-6354, USA}
\footnote{$^{}$Computer Science and Mathematics Division, Oak Ridge National Laboratory, P.O.Box 2008, Oak Ridge, TN 37831-6164, USA}
\footnote{$^{}$Department of Physics, Embry Riddle Aeronautical University, Prescott, AZ 86301-3720,USA}
\footnote{$^{}$Physics Division, National Science Foundation, Arlington, VA 22207 USA}
\footnote{$^{}$Joint Institute for Nuclear Physics and its Applications, Oak Ridge National Laboratory, P.O. Box 2008, Oak Ridge, TN 37831-6374, USA}
\footnote{$^{}$Department of Physics, Florida Atlantic University, 777 Glades Road, Boca Raton, FL 33431-0991, USA}
\footnote{$^{}$National Center for Computational Sciences, Oak Ridge National Laboratory, P.O. Box 2008, Oak Ridge, TN 37831-6164, USA}
\footnote{$^{}$Department of Physics, North Carolina State University,  Raleigh, NC 27695-8202, USA}
\footnote{$^{}$Lawrence Berkeley National Lab, Berkeley, CA 94720, USA
}

\maketitle\abstracts{
In this work we report briefly on the gravitational wave (GW) signal computed in the context of a self-consistent, three-dimensional (3D) simulation of a core-collapse supernova (CCSN) explosion of a 15~\msun progenitor star. 
We present a short overview of the GW signal, including signal amplitude, frequency distribution, and the energy emitted in the form of GWs for each phase of explosion, along with neutrino luminosities, and discuss correlations between them.}

\section{Introduction}

The era of gravitational wave (GW) astronomy began with the first direct detections of GW signals from binary-black-hole mergers~\cite{gw1}$^,\,$\cite{gw2}.  
Among the sources of Earth-detectable GWs we can include core-collapse supernovae (CCSNe), which are physics rich.  
They result from the rapid collapse and violent explosion of a massive star (\emph{M}$>$8\msun), and are one of the most energetic burst phenomena in the Universe.  
An enormous amount of energy ($\sim 10^{53}$~erg) is released in the form of neutrinos, synthesized matter, and electromagnetic and gravitational radiation.  
Simultaneous detection of these signals will help us reveal details of the physical processes taking place under the extreme conditions as a massive star explodes [for a comprehensive review, see, e.g., Ott (2009)~\cite{ott}].  
Thus, any advanced study of CCSN requires detailed simulation of the processes in the supernova core, and physical fidelity of CCSN simulations is crucial.  
However, self-consistent CCSN simulations that include the physics necessary for production of realistic GW templates (e.g., neutrino transport with a complete set of weak interactions and effects of general relativity) are extremely compute intensive, and demand computing at the extreme scale.  
For instance, the number of processors required to perform the 3D simulation presented here was on the order of $10^5$, and the full simulation, until the onset of explosion (350--400~ms) required $\sim10^8$ CPU hours.  
In spite of recent advances in simulation codes and computational resources, few groups~\cite{le}$^,\,$\cite{an}$^,\,$\cite{ku} have performed 3D simulations that include the necessary physics for realistic GW templates.  
Since CCSNe are one of the most promising sources for multimessenger astronomy due to their strong combined neutrino and GW signals, these 3D models are able to produce the most reliable GW and neutrino signals, and deduce potential correlations between them, and could help not only detect GW signals from CCSN, but also help resolve a number of open questions concerning the physics of the CCSN explosion mechanism.  
To contribute to these efforts, we present here a short overview of GW and neutrino signals from a 3D simulation carried out with the \chimera~code.

\section{Model Description}

\begin{wrapfigure}{R}{0.5\textwidth}
\centering
\includegraphics[width=0.45\textwidth]{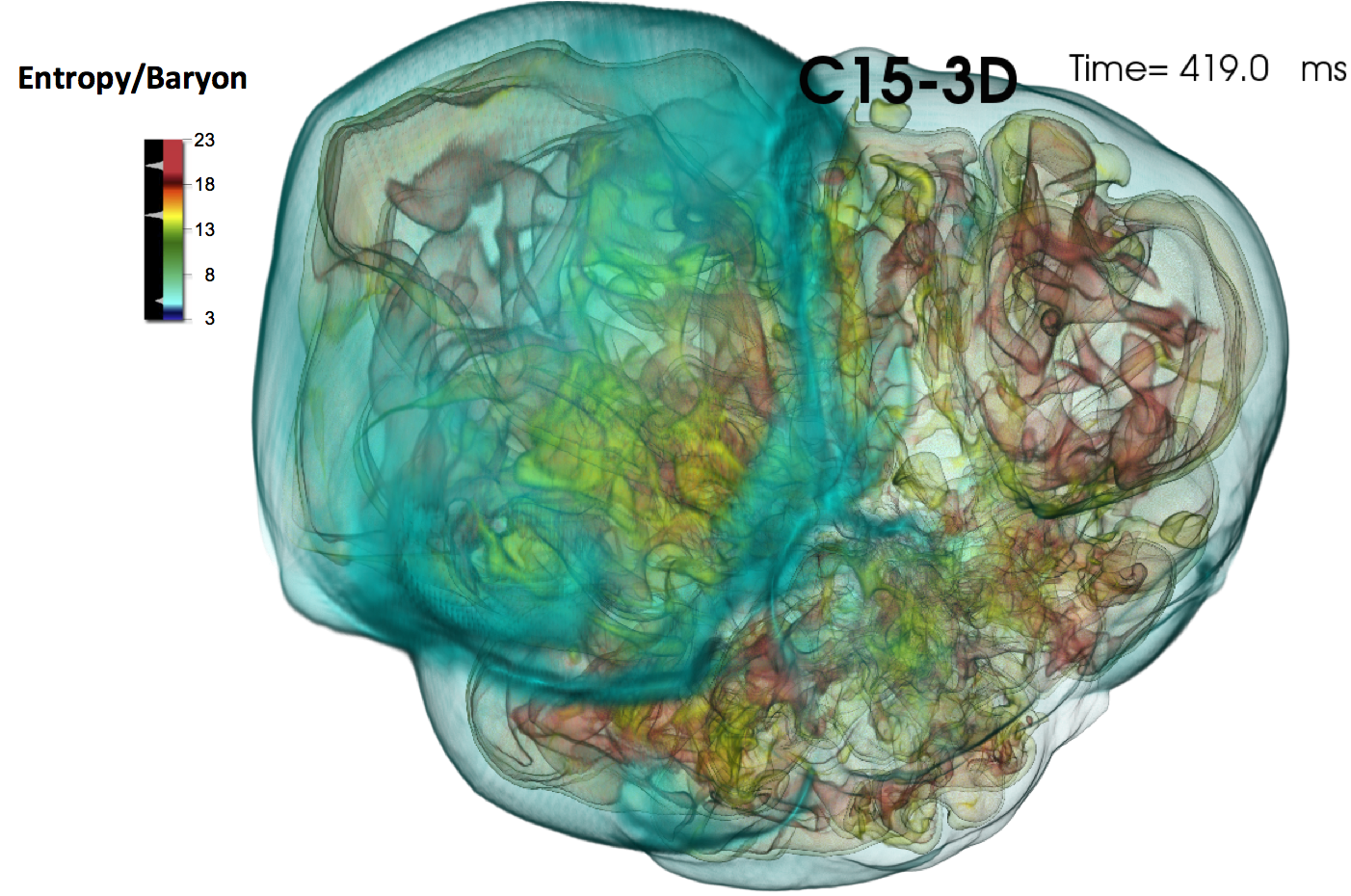}
\caption{\label{fig:snapshot} Volume rendering of the specific entropy from a CCSN explosion simulation with the \chimera~code. 
The large rising plumes drive the expansion of the shock, while active accretion onto the PNS surface sustains neutrino emission, and contributes to the GW signal.}
\end{wrapfigure}

Our analysis is based on the 3D CCSN simulation of the 15\msun progenitor of Woosley \& Heger~\cite{wo}, performed by Lentz\etal~\cite{le} (Fig.~\ref{fig:snapshot}).  
The simulation is performed with the neutrino radiation hydrodynamics code \chimera\, that includes neutrino transport with a modern set of weak interactions through the "ray-by-ray-plus" approach, a nuclear equation of state, and an effective relativistic gravitational potential.  
The effective gravitational potential for self-gravitating fluids is build from a Newtonian potential by replacing the leading term in its spherical harmonic expansion with the TOV potential.  
The effective potential mimics the deeper potential well obtained in full GR~\cite{ma}.  
Gravitational redshift of the neutrino spectrum is also included in the transport scheme.  
The 3D computational grid consist of 540($r$)$\times$180($\theta$)$\times$180($\phi$).  
The $\theta$-resolution varies from $ 2/3^\circ$ near the equator to $8.5^\circ$ near the poles.  
The $\phi$-resolution is uniform and equal to $2^\circ$.  
The radial resolution varied according to the structure of the collapsed core in our adaptive mesh, and reached $\sim$0.1~km inside the PNS. 

\section{Results}

A CCSN passes through several phases before the onset of explosion.  
The massive star's iron core collapses, with inner core densities ultimately exceeding nuclear densities.  
At these extremes, the inner core rebounds due to the stiffening of the EoS and launches an outgoing shock wave.  
Within a $\sim100$~ms after bounce, the shock stalls due to energy loss by neutrino emission and dissociation of iron falling through the shock.  
It is revived by neutrino energy deposition behind the shock, but aided by hydrodynamical instabilities~\cite{ja}.  
Instabilities operating inside the proto-neutron star (PNS), large scale convection in the neutrino heated, post-shock region, and the standing accretion shock instability (SASI), contribute to produce a strong GW burst signal that can be detected by GW observatories (cf. Fig.~\ref{fig:snapshot}).  

\subsection{GW signal}

\begin{figure}
\begin{minipage}{0.5\linewidth}
\centerline{\includegraphics[width=1.1\linewidth]{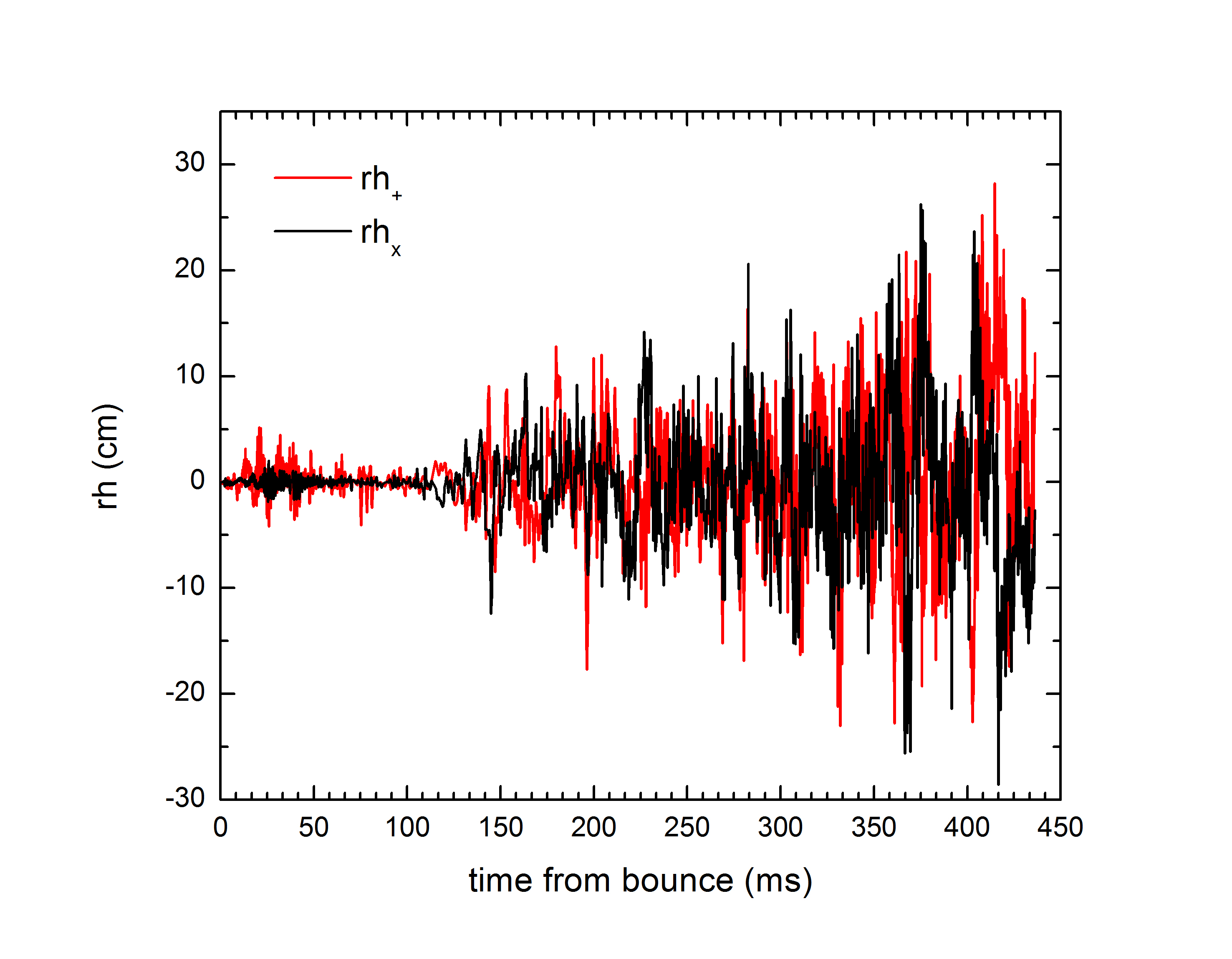}}
\end{minipage}
\hfill
\begin{minipage}{0.5\linewidth}
\centerline{\includegraphics[width=0.9\linewidth]{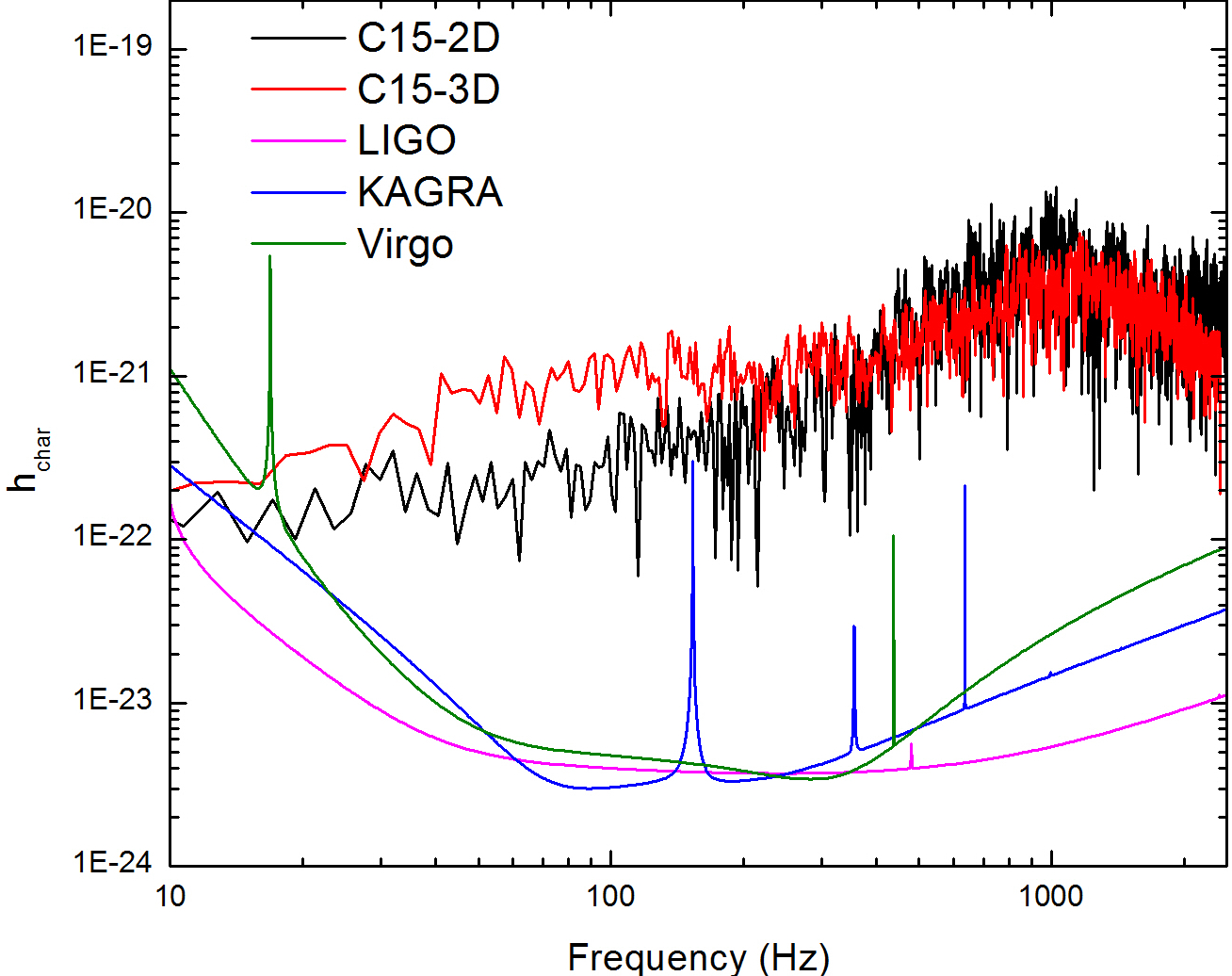}}
\end{minipage}
\vfill
\begin{minipage}{0.5\linewidth}
\centerline{\includegraphics[width=0.9\linewidth]{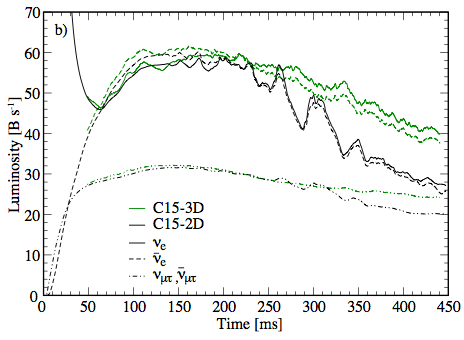}}
\end{minipage}
\hfill
\begin{minipage}{0.5\linewidth}
\centerline{\includegraphics[width=1.1\linewidth]{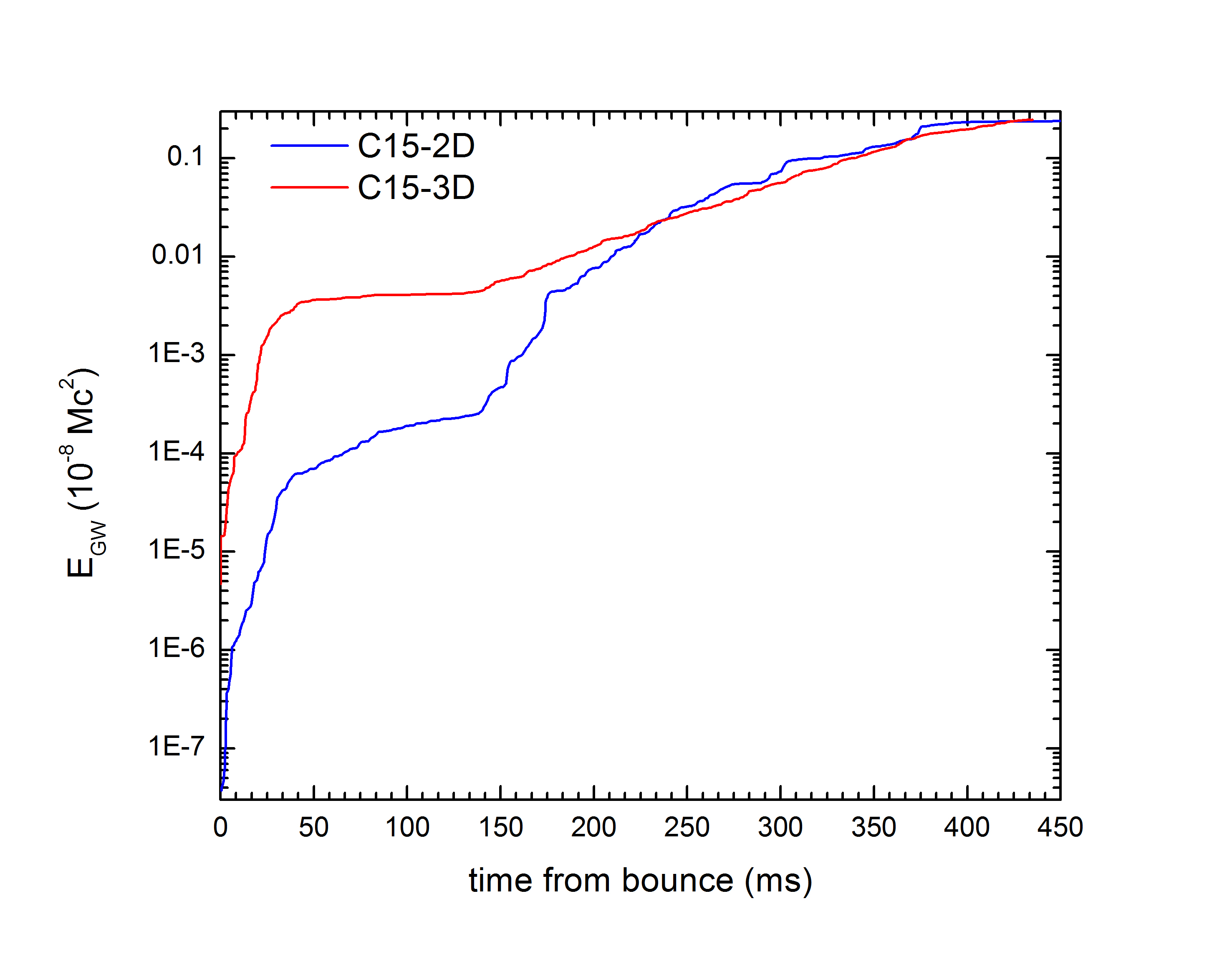}}
\end{minipage}
\caption[]{Two polarizations of the GW signal as they are seen by observer at the equator (top left), characteristic spectral strain $h_\textrm{char}$ of 2D and 3D models at a distance of 10~kpc compared with the design noise levels  of the current and future GW detectors (top right), neutrino luminosities for each type of neutrinos using in the simulation (bottom left), energy emitted in the form of GWs for C15-2D and C15-3D models (bottom right).  }
\label{fig:gw}
\end{figure}

The typical GW signal from axisymmetric (2D) explosion models, carried out to about two seconds after bounce, consists of four main phases~\cite{ya15}.  
The structure of our 3D waveform is similar to what was obtained in the 2D case.  
The early GW signal is produced by prompt convection inside the PNS that develops behind the quickly expanding, newly formed shock. 
It lasts for 70--80~ms after bounce.  
The GW energy emitted during the early phase is $\sim4\times10^{-11}$\msun c$^2$ (Fig.~\ref{fig:gw} bottom right).  
Note that the GW energy of the early signal is much higher in 3D than in 2D.  
This can be explained by the inverse turbulent cascade in 2D:  turbulent energy accumulates artificially in large scale eddies, decreasing the emitted GW energy.  
The quiescent phase (80--120~ms) is the result of shock stagnation and ceasing of the initial prompt convection.  
Other fluid instabilities have not yet developed.  
The strongest phase of the GW signal starts at around 120~ms.  
It is the result of a non-linear interaction between different types of instabilities. Active accretion on the PNS surface induces the density disturbances inside the PNS (g-modes) that produce the most energetic GW signal. In turn, active accretion is formed due to the superposition of neutrino-driven convection and the SASI.  
The GW energy emitted during the strong signal phase is $\sim2\times10^{-9}$\msun c$^2$, which is almost two orders of magnitude higher than the GW energy emitted in the early phase.  
In contrast with the early phase, more GW energy is emitted in the 2D model than in the 3D model during the strong phase.  
Here, the additional dimension allows development of global, low frequency modes (e.g., spiral SASI modes and large scale convection) that are suppressed in the 2D models.  
The spectrogram of the GW signal confirms the presence of stronger low-frequency (50--200 Hz) activity in the signal (Fig.~\ref{fig:gw}; top right).  
The fourth phase -- DC offset (``tail") in the signal observed in the 2D models due to shock expansion~\cite{ya15} is not yet present in the 3D model.  
The onset of explosion in the 3D model is delayed by about 100~ms relative to the 2D model~\cite{le}.  
Consequently, during the 450~ms considered here, we do not observe the low-frequency tail in the GW signal.  
It may appear later, but will probably be less pronounced due to the different explosion morphologies in 2D and 3D.  
In the 2D models, the heating region is divided into 2--3 large volumes due to the presence of 1--2 accretion downflows. When these volumes acquire enough energy from neutrinos they begin to expand at the same time. The picture is different in the 3D model. There are multiple down flows splitting the heating region into multiple volumes of relatively small sizes. When one of these volumes becomes sufficiently large it starts to expand while others do not~\cite{le} (Fig.~\ref{fig:snapshot}). 

\subsection{Neutrino signal}

Neutrinos play a fundamental role in CCSNe.  
They carry most (99\%) of the gravitational binding energy liberated during iron core collapse.  
As such, neutrinos are the major source of energy for the supernova explosion in the delayed neutrino-driven mechanism.~\cite{ja}.  
Moreover, the observation of supernova neutrinos from a CCSN will provide not only important insights into the dynamics inside the exploding star, but will also significantly increase the chances for detection of the GW signal from the supernova event.  

Neutrino luminosities from the 3D model are plotted in the bottom left panel in Fig.~\ref{fig:gw}.  
The first prominent feature is the well-known electron neutrino, or neutronization, burst that occurs when the post-shock matter becomes transparent to neutrinos shortly after bounce.  
During the later phase, the electron neutrino and antineutrino luminosities are largely regulated by the mass accretion onto the PNS.  
(The neutrino luminosities are also sensitive to PNS parameters; e.g., compactness, surface temperature.)   
As the accretion rate drops, the relative contribution of the electron neutrino and antineutrino flux decreases.  
The significant variations in the neutrino luminosity for $t>150$ is due to variations in mass accretion (in response to fluid instabilities operating).  
The variations in the $\nu$-luminosity in 2D are larger than in 3D because multiple accretion down flows in the 3D model are less massive than the 1--2 down flows present in the 2D model.  
The heavy flavor neutrinos ($\nu_{\mu},\nu_{\tau}$; emanating deeper in the PNS) show a steady decline after about 150~ms, and are not as sensitive to sudden variations in accretion.  
It is worth noting that the fluctuations of the $\nu$-luminosities are a consequence of oscillations in mean shock radius.  
Thus, the SASI activity is imprinted on neutrino signal.  
When comparing the top and bottom left panels of Fig.~\ref{fig:gw}, a correlation between the neutrino and GW signals can be inferred.  
Immediately after the neutronization burst, there are no significant variations in the neutrino luminosities.  
This phase coincides with the quiescent phase of the GW signal.  
The strongest variations in the neutrino signal coincide with the large amplitude spikes in the GW signal.  
This is to be expected since both features are due to variations in accretion onto the PNS, and a response to explosion dynamics.   

\section{Outlook}

Despite the complexity of supernova physics, several groups are now able to simulate CCSN explosions with high physical fidelity, and compute GW and neutrino signals from 3D models~\cite{ya17}$^,\,$\cite{an}$^,\,$\cite{ku}.  
However, only a handful of models exist today. 
The 3D simulations of Lentz\etal~\cite{le} and  Andresen\etal~\cite{an} have similar physics input: spectral neutrino transport with full set of weak interactions and effective GR potential. The fully relativistic simulation of Kuroda\etal~\cite{ku} includes grey neutrino transport and a reduced set of weak interactions.  
Each of these groups still employs approximations to be able to complete simulations, but work on improvements are underway; e.g., better resolution, better algorithms well suited to modern computer architectures, and improved physical fidelity in both the gravity and the neutrino transport sectors.  
There are robust features of the GW signals that are common across simulations, but there are differences in the waveforms too~\cite{ya17}$^,\,$\cite{an}.  
These differences motivate us to establish closer collaborations between supernova modeling efforts and to look deeper at the problem.  
By continuing the work to model CCSN, with independent codes, we will obtain a larger catalog and more robust waveforms to aid in GW searches.  

\section*{Acknowledgments}

This research was supported by the U.S. Department of Energy Offices of Nuclear Physics and Advanced Scientific Computing Research, the NASA Astrophysics Theory and Fundamental Physics Program (grants NNH08AH71I and NNH11AQ72I), and the National Science Foundation PetaApps Program (grants OCI-0749242, OCI-0749204, and OCI-0749248) and Gravitational Physics Program (grant GP-1505933).

\end{document}